\documentclass[a4paper,11pt]{article}
\usepackage{jheppub} 
\usepackage{amsmath}
\usepackage{xspace}

\usepackage{tabularx}
\usepackage{tikz}
\usepackage[compat=1.1.0]{tikz-feynhand}
\usetikzlibrary{patterns, math}
\usetikzlibrary{positioning}

\usepackage{booktabs}

\title{Prospects for studying the $WH\gamma$ process in $pp$ collisions at the LHC}

\newcommand{\MG}{\texttt{MadGraph\_aMC\@NLO} }
\newcommand{\Pythia}{\texttt{Pythia 8.3} }
\newcommand{\Delphes}{\texttt{Delphes 3.5} }
\newcommand{\BDT}{\texttt{XGBoost} }
\newcommand{\pT}{\ensuremath{p_\mathrm{T}}\xspace}
\newcommand{\WHG}{\ensuremath{WH\gamma}\xspace}
\newcommand{\GeV}{\ensuremath{\,\text{Ge\hspace{-.08em}V}}\xspace}
\newcommand{\pTmiss}{\ensuremath{\pT^\text{miss}}\xspace}
\newcommand{\pTlep}{\ensuremath{\pT^{\ell}}\xspace}
\newcommand{\mt}{\ensuremath{m_\mathrm{T}}\xspace}

\author{Youpeng Wu,}
\emailAdd{youpeng@stu.pku.edu.cn}
\affiliation{State Key Laboratory of Nuclear Physics and Technology, School of Physics, Peking University, Beijing, 100871, China}
\author{Jie Xiao,}
\affiliation{Institute for Particle Physics Phenomenology, Durham University, Durham DH1 3LE, UK}
\emailAdd{jie.xiao@durham.ac.uk}
\author{Andrew Michael Levin,}
\emailAdd{andrew.michael.levin@cern.ch}
\affiliation{State Key Laboratory of Nuclear Physics and Technology, School of Physics, Peking University, Beijing, 100871, China}
\author{Qiang Li}
\emailAdd{qliphy0@pku.edu.cn}
\affiliation{State Key Laboratory of Nuclear Physics and Technology, School of Physics, Peking University, Beijing, 100871, China}

\abstract{
The Standard Model of particle physics, though remarkably successful, leaves open several major questions that continue to motivate searches for new phenomena. Multiboson interactions involving the Higgs boson are of special interest as probes of the electroweak Lagrangian where potential new physics may be hiding. In this work, we present a study of the simultaneous production of a W boson, a Higgs bosons and a photon in proton-proton collisions at the Large Hadron Collider. Monte Carlo simulation is performed to model both the signal and the background processes, and detector effects are included according to CMS specifications. Boosted decision trees are employed to optimize the event selection and enhance signal-background discrimination. We estimate that with an integrated luminosity of 440~$\rm fb^{-1}$, the expected significance for the \WHG process is 0.63$\sigma$, projected to reach 1.64$\sigma$ at the High-Luminosity LHC (HL-LHC).
}

\begin{document}
\maketitle
\flushbottom

\section{Introduction}

The Standard Model (SM) of particle physics is a well-established theoretical framework that describes the fundamental particles and their interactions~\cite{gaillard1999standard}, and the discovery of the Higgs boson at the CERN Large Hadron Collider (LHC) has provided a fact that SM is nearly complete~\cite{aad2012observation,chatrchyan2012observation}. However, complete dosen't mean perfect, there are still many open questions in particle physics that the SM cannot answer. Studying processes involving the Higgs boson is crucial for testing the SM and searching for new physics beyond it. 
Currently there are many studies on di-boson final states containing Higgs boson~\cite{godbole2003search,hayrapetyan2025study,hernandez2021anomalous,cms2022measurement}, 
but few researches have been conducted on three bosons final state~\cite{belforte2014search,cms2019search,teles2013search,mozer2016nonresonant,atlas2025constraining}, and handful of them involve Higgs boson in previous studies ~\cite{cms2025measurement,hayrapetyan2025measurement,cappati2022sensitivity}.
Therefore, it currently appears necessary to study multi-boson final states containing Higgs bosons. The \WHG process, as a rare three boson production channel involving the W boson, Higgs boson, and a photon, offers a unique probe into the electroweak part of the SM. Additionally, it provides insights into Higgs rare decays, such as \(H\to b\bar{b}\gamma\), which are suppressed in the SM but could be enhanced in beyond-SM scenarios, making \WHG an ideal channel for precision measurements and rare event searches~\cite{han2017radiative,jakobs2024profile}.

The LHC (Large Hadron Collider) provides an excellent opportunity to study processes involving the Higgs boson due to its high energy and luminosity~\cite{evans2008lhc} . 
While the Higgs boson has been extensively studied in di-boson channels like ZZ and WW, previous studies on the \WHG process remain limited, primarily due to its low cross-section and challenging experimental signatures. 
However, with the increased luminosity from LHC Run 3 and the upcoming High-Luminosity LHC (HL-LHC), more precise measurements are expected, enabling deeper investigations into this rare process~\cite{apollinari2017high,fartoukh2021lhc}. The CMS detector, with its advanced tracking, calorimetry, and muon systems, is particularly well-suited for identifying the leptonic decays of the W boson, the di-jet signature of the Higgs, and the isolated photon in \WHG events.

\begin{figure}[htb]
    \centering
    \scalebox{0.65}{
\begin{tikzpicture}
    \begin{feynhand}
        \vertex (f) at (-2,2) {\(f\)};
        \vertex (fbar) at (-2,-2) {\(\bar{f}\)};
        \vertex (ffw) at (-0.5,0);
        \vertex (wwh) at (0.5,0);
        \vertex [NWblob] (hbb) at (2,1) {};
        \vertex (b1) at (4,2) {\(b\)};
        \vertex (b2) at (4,0) {\(\bar{b}\)};
        \vertex (g) at (4,1) {\(\gamma\)};
        \vertex (w1) at (2,-1.5);
        \vertex (l1) at (4,-1) {\(\ell\)};
        \vertex (l2) at (4,-2) {\(\bar{\nu}\)};

        \propag[fer] (f) to (ffw);
        \propag[fer] (ffw) to (fbar);
        \propag[bos] (ffw) to [edge label=\(W\)](wwh);
        \propag[bos] (wwh) to [edge label=\(H\)](hbb);
        \propag[bos] (wwh) to [edge label=\(W\)](w1);
        \propag[fer] (hbb) to (b1);
        \propag[antfer] (hbb) to (b2);
        \propag[pho] (hbb) to (g);
        \propag[fer] (w1) to (l1);
        \propag[antfer] (w1) to (l2);
    \end{feynhand}
\end{tikzpicture}
}
\hspace*{1em}
\scalebox{0.65}{
\begin{tikzpicture}
    \begin{feynhand}
        \vertex (f) at (-2,2) {\(f\)};
        \vertex (fbar) at (-2,-2) {\(\bar{f}\)};
        \vertex (ffw) at (-0.5,0);
        \vertex (wwh) at (0.5,0);
        \vertex (hbb) at (2,1) ;
        \vertex (bgb) at (3,1.5) ;
        \vertex (b1) at (4,2) {\(b\)};
        \vertex (b2) at (4,0) {\(\bar{b}\)};
        \vertex (g) at (4,1) {\(\gamma\)};
        \vertex (w1) at (2,-1.5);
        \vertex (l1) at (4,-1) {\(\ell\)};
        \vertex (l2) at (4,-2) {\(\bar{\nu}\)};

        \propag[fer] (f) to (ffw);
        \propag[fer] (ffw) to (fbar);
        \propag[bos] (ffw) to [edge label=\(W\)](wwh);
        \propag[bos] (wwh) to [edge label=\(H\)](hbb);
        \propag[bos] (wwh) to [edge label=\(W\)](w1);
        \propag[pho] (bgb) to (g);
        \propag[fer] (hbb) to (b1);
        \propag[antfer] (hbb) to (b2);
        \propag[fer] (w1) to (l1);
        \propag[antfer] (w1) to (l2);
    \end{feynhand}
\end{tikzpicture}
}
\hspace*{1em}
\scalebox{0.65}{
\begin{tikzpicture}
    \begin{feynhand}
        \vertex (f) at (-2,2) {\(f\)};
        \vertex (fbar) at (-2,-2) {\(\bar{f}\)};
        \vertex (ffw) at (-0.5,0);
        \vertex (wwh) at (0.5,0);
        \vertex (hzz) at (2,1) ;
        \vertex (zzg) at (3.5,1.5) ;
        \vertex (zzv) at (3.5,0.5) ;
        \vertex (vbb) at (4,0.5) ;
        \vertex (b1) at (5,1) {\(b\)};
        \vertex (b2) at (5,0) {\(\bar{b}\)};
        \vertex (g) at (5,2) {\(\gamma\)};
        \vertex (w1) at (2,-1.5);
        \vertex (l1) at (4,-1) {\(\ell\)};
        \vertex (l2) at (4,-2) {\(\bar{\nu}\)};

        \propag[fer] (f) to (ffw);
        \propag[fer] (ffw) to (fbar);
        \propag[bos] (ffw) to [edge label=\(W\)](wwh);
        \propag[bos] (wwh) to [edge label=\(H\)](hzz);
        \propag[bos] (wwh) to [edge label=\(W\)](w1);
        \propag[fer] (vbb) to (b1);
        \propag[antfer] (vbb) to (b2);
        \propag[bos] (zzv) to (vbb);
        \propag[pho] (zzg) to (g);
        \propag[fer] (hzz) to (zzv);
        \propag[antfer] (hzz) to  [edge label=\(Z\)] (zzg);
        \propag[fer] (zzv) to (zzg);
        \propag[fer] (w1) to (l1);
        \propag[antfer] (w1) to (l2);
    \end{feynhand}
\end{tikzpicture}
}
\newline 
\scalebox{0.65}{
\begin{tikzpicture}
    \begin{feynhand}
        \vertex (f) at (-2,2) {\(f\)};
        \vertex (fbar) at (-2,-2) {\(\bar{f}\)};
        \vertex (ffw) at (-0.5,0);
        \vertex (wwh) at (0.5,0);
        \vertex (hbb) at (2,1.5);
        \vertex (b1) at (4,2) {\(b\)};
        \vertex (b2) at (4,1) {\(\bar{b}\)};
        \vertex (w1) at (2,-1.5);
        \vertex (llg) at (3,-0.5);
        \vertex (l1) at (4,0.5) {\(\ell\)};
        \vertex (g) at (4,-1) {\(\gamma\)};
        \vertex (l2) at (4,-2) {\(\bar{\nu}\)};

        \propag[fer] (f) to (ffw);
        \propag[fer] (ffw) to (fbar);
        \propag[bos] (ffw) to [edge label=\(W\)](wwh);
        \propag[bos] (wwh) to [edge label=\(H\)](hbb);
        \propag[bos] (wwh) to [edge label=\(W\)](w1);
        \propag[fer] (hbb) to (b1);
        \propag[antfer] (hbb) to (b2);
        \propag[fer] (w1) to (llg);
        \propag[fer] (llg) to (l1);
        \propag[pho] (llg) to (g);
        \propag[antfer] (w1) to (l2);
    \end{feynhand}
\end{tikzpicture}
}
\hspace*{1em}
\scalebox{0.65}{
\begin{tikzpicture}
    \begin{feynhand}
        \vertex (f) at (-2,2) {\(f\)};
        \vertex (fbar) at (-2,-2) {\(\bar{f}\)};
        \vertex (ffw) at (-0.5,1);
        \vertex (ffg) at (-0.5,-1);
        \vertex (g) at (4,-2) {\(\gamma\)};
        \vertex (wwh) at (1,1);
        \vertex (hbb) at (2,1.5);
        \vertex (b1) at (4,2) {\(b\)};
        \vertex (b2) at (4,1) {\(\bar{b}\)};
        \vertex (w1) at (2,0);
        \vertex (l1) at (4,-0) {\(\ell\)};
        \vertex (l2) at (4,-1) {\(\bar{\nu}\)};

        \propag[fer] (f) to (ffw);
        \propag[fer] (ffw) to [edge label=\(\bar{f}\)](ffg);
        \propag[bos] (ffw) to [edge label=\(W\)](wwh);
        \propag[bos] (wwh) to [edge label=\(H\)](hbb);
        \propag[bos] (wwh) to [edge label=\(W\)](w1);
        \propag[fer] (hbb) to (b1);
        \propag[antfer] (hbb) to (b2);
        \propag[fer] (w1) to (l1);
        \propag[antfer] (w1) to (l2);
        \propag[pho] (ffg) to (g);
        \propag[antfer] (fbar) to (ffg);
    \end{feynhand}
\end{tikzpicture}
}
    \caption{Feynman diagram for \WHG signal process. Left-above: Higgs radiation, Middle-above: QED process,  Right-above: \(H\to Z\gamma\) process; Left-bottom: W boson decay (leptonic), Right-bottom: quark radiation. And we consider W boson decay and quark radiation as non-Higgs process.}
    \label{fig:signalprocess}
\end{figure}
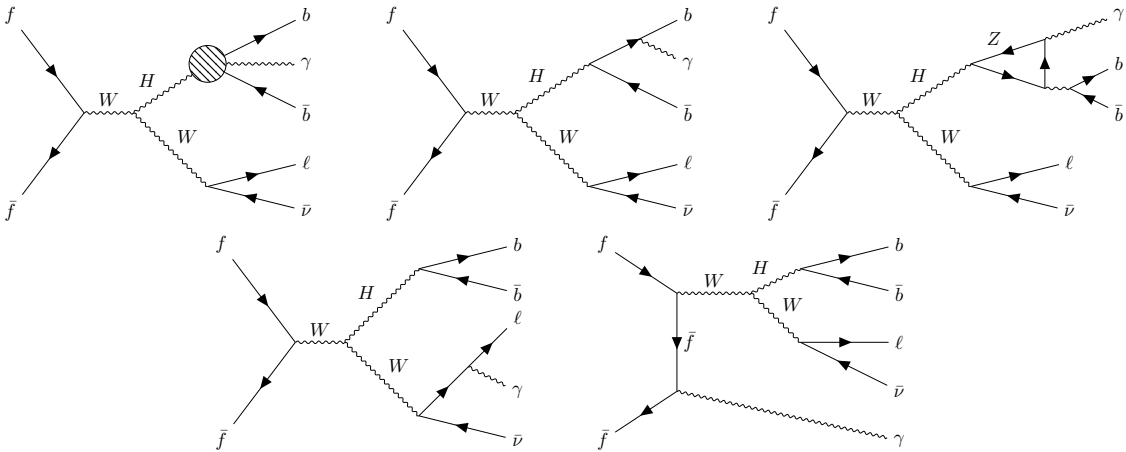

In this study, \WHG include process
defined as figure \ref{fig:signalprocess}, where the photon is radiated either from the Higgs boson (Higgs process) or from the W boson or quark (non-Higgs process).
Both them contribute to the same final state structure of one lepton (electron or muon), missing transverse energy (from neutrino), two b-jets (from Higgs decay) and one photon. We focus on the leptonic decay channel of the W boson due to its cleaner experimental signature compared to hadronic decays, which are often plagued by large QCD backgrounds while we consider Higgs decay to \(b\bar{b}\gamma\) or \(b\bar{b}\) final states. We use Monte Carlo (MC) simulation tools include \MG ~\cite{mg5alwall2014automated}, \Pythia ~\cite{py8bierlich2022comprehensive} and \Delphes ~\cite{de2014delphes} to generate and simulate both signal and background events. To enhance the precision of our analysis, we use a efficiency field theory model implemented in UFO format to generate the \(H\to b\bar{b} \gamma\) decay process by MadSpin~\cite{artoisenet2013automatic} after generating the \WHG process without Higgs decay. 
This enables keeping the dual-peak feature of the photon energy spectrum in the study ~\cite{han2017radiative}, provided by $QED$ processes and $EW+\gamma$ processes respectively.
We also employ machine learning techniques, specifically Boosted Decision Trees (BDT) using the \BDT package~\cite{chen2016xgboost}, to optimize event selection and improve signal-background discrimination. This approach allows us to effectively identify the rare \WHG events amidst significant SM backgrounds.

\section{Event Reconstruction and Selection}

We categorize the signal process into three distinct parts: Higgs boson, W boson, and quark, based on the origin of the photon produced in each case. 

\begin{figure}[htb]
    \centering
    \includegraphics[width=0.5\textwidth]{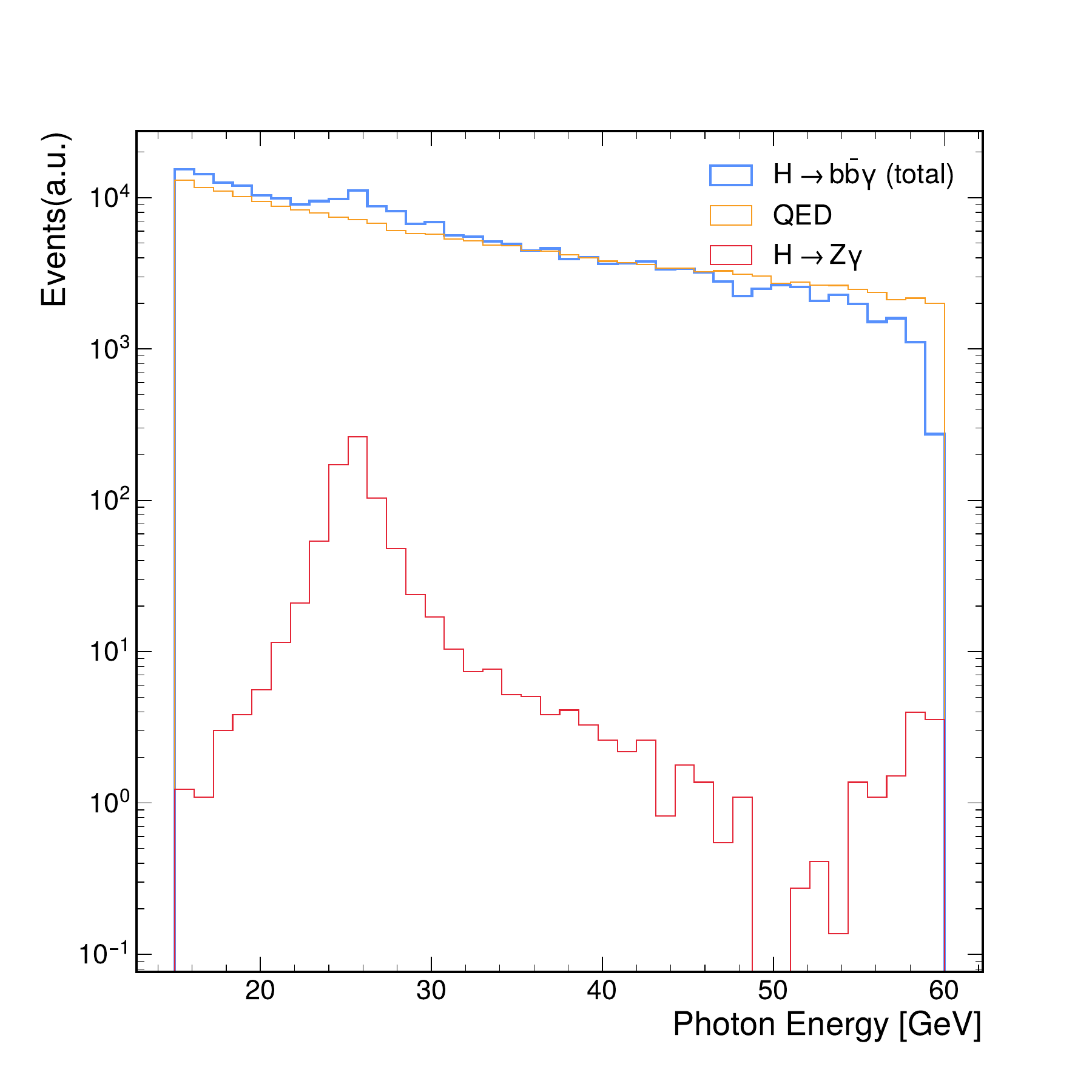}
    \caption{Photon energy distribution for WHG Higgs process at 13 TeV. The QED process is processed by photon radiation from b quark, while the EW+\(\gamma\) process is processed by electroweak one-loop correction with photon radiation from Higgs boson.}
    \label{fig:photonE} 
\end{figure}

In our study, we divide the production process into two parts. One part corresponds to the Higgs process, which is generated using the UFO model. We first generate samples when the Higgs boson has not yet decayed, then use MadSpin combined with UFO model to make the Higgs decay to $bb\gamma$. It contain two photon radiation process, one is normal QED process where photon is radiated from b quark, the other is electroweak one-loop correction with photon radiation from Higgs boson which is contributed by \(h\to \gamma Z^{\star} \to f\bar f\gamma\), \(h\to \gamma \gamma^{\star} \to f\bar f\gamma\) and else. The photon energy distribution for both processes is shown in Fig.~\ref{fig:photonE} with previous research~\cite{han2017radiative}. 

The other part is the non-Higgs process, for which we employ the \textit{sm} model for generation. For both processes, we use leading order (LO) level using \MG. When calculating the cross section for both the signal and background process, we use the default Parton Distribution Functions (PDFs) set NNPDF23 and set the factorization and normalization scale to the physical mass of boson~\cite{ball2013parton}. 

Simulated events for the irreducible background processes containing four prompt leptons in the final state, such as \(t\bar t\gamma \), \(tW\gamma/\bar{t}W\gamma\), \(W/Z+\gamma+\rm Jets\) and \(WW\gamma/W Z\gamma\)
are generated with \MG as for the possible background process. The background processes considered in this analysis are normalized using their production cross sections \ref{tab:background}. The these events are passed to \Pythia for further event hadronization. Eventually all events are passed though \Delphes program to include detector effects with a configured file designed for CMS detector and as a preliminary particle object reconstruction. After \Delphes reconstruction, we get the \texttt{Photon}, \texttt{Electron} and \texttt{Muon} object.

\begin{table}
    \centering
    \caption{Cross sections for background processes at 13 TeV}
    \label{tab:background}
    \begin{tabularx}{\textwidth}{l|XX}
        \toprule
        Process & Cross Section (pb) & Generate Events \\
        \midrule
        \(p\ p \to t\bar{t}\gamma\) & 2.703 \(\pm\) 0.009074 & 1,000,000 \\
        \(p\ p \to t\bar{t}\) & 214.1 \(\pm\) 0.036 & 3,000,000 \\
        \(p\ p \to tW\gamma\) & 0.315 \(\pm\) 0.00016 &  100,000\\
        \(p\ p \to W/Z+\gamma+\rm Jets\) & 116.2 \(\pm\) 0.3538 & 100,000  \\
        \(p\ p \to WW\gamma/W Z\gamma\) & 0.3564 \(\pm\) 0.0009274 & 100,000 \\
        \bottomrule
    \end{tabularx}
\end{table}

Electron are reconstructed from quick detector simulation presented by \Delphes. The electron tracking efficiency and efficiency formula is dependent on the transverse momentum \pT and pseudorapidity \(\eta\) of the electron. The electron momentum smear function is dependent on the research of CMS collaboration~\cite{cms2015performance}. Electron isolation requirement is based on the scalar sum of the transverse momenta of all particle flow candidates in a cone of size \(\Delta R = 0.5\) around the electron direction, excluding the electron itself. Muon candidates are reconstructed by efficiency formula and momentum smearing dependent on \pT and \(\eta\) similar to electron. 

Jet from parton showering and hadronization are reconstructed using the anti-\(k_{t}\) algorithm with a \(\Delta R\) less than 0.5 and minimum \pT of 20\GeV. BTagging is performed by \Delphes with an efficiency presented from light flavor jets and c-jets mistag rate dependent on jet \pT and \(\eta\), while the b-jet tagging efficiency is set to a formula dependent on jet \pT and \(\eta\)~\cite{cms2013identification}. 

Photon candidates are reconstructed from \Delphes which simulated the CMS detector ECal response. Assume \(0.02\times 0.02\) resolution in \(\eta-\phi\) in the barrel \(|\eta|<1.5\) and \(1.5 < |\eta| < 3.0\) for HGCAL-ECAL. Set ECal resolution formula as \(\eta\) shape and energy shape from research ~\cite{cms2013energy,cms2015performance}. Photon isolation is determined by summing the transverse momenta of all particle flow candidates within a cone of \(\Delta R = 0.5\) centered on the photon, omitting the photon itself from the calculation.

In order to select the events of interest, we apply a series of selection criteria. To select WHG signal events, we require a good lepton which is from leptonic W boson decay. Either on muon \((\pT > 26\GeV, |\eta|< 2.4)\) or one electron \((\pT > 30 \GeV, |\eta| < 2.4)\) is required as one lepton trigger. Events are also required to have \pTmiss greater than 30\GeV to ensure neutrino produced in event. In addition, both \(H\to b\ \bar{b}\ \gamma\) and \(H \to b\ \bar{b}\) decay model need two b quark, so we need event containing at least 2 b-tagging jet to simulate two AK4 jets in CMS. Because both b-jet decay from one higgs boson, we require that \(\Delta R_{j_1j_2}\) less than 4, and angle of direction are close that difference in pseudorapidity between the jets \(\Delta \eta_{j_1j_2} < 3.0\) and angle \(\Delta \phi_{j_1j_2} < 2.0\). We also require a photon with large \(\pT > 20\GeV\).

\begin{figure}[htb]
    \centering
    \includegraphics[width=0.48\textwidth]{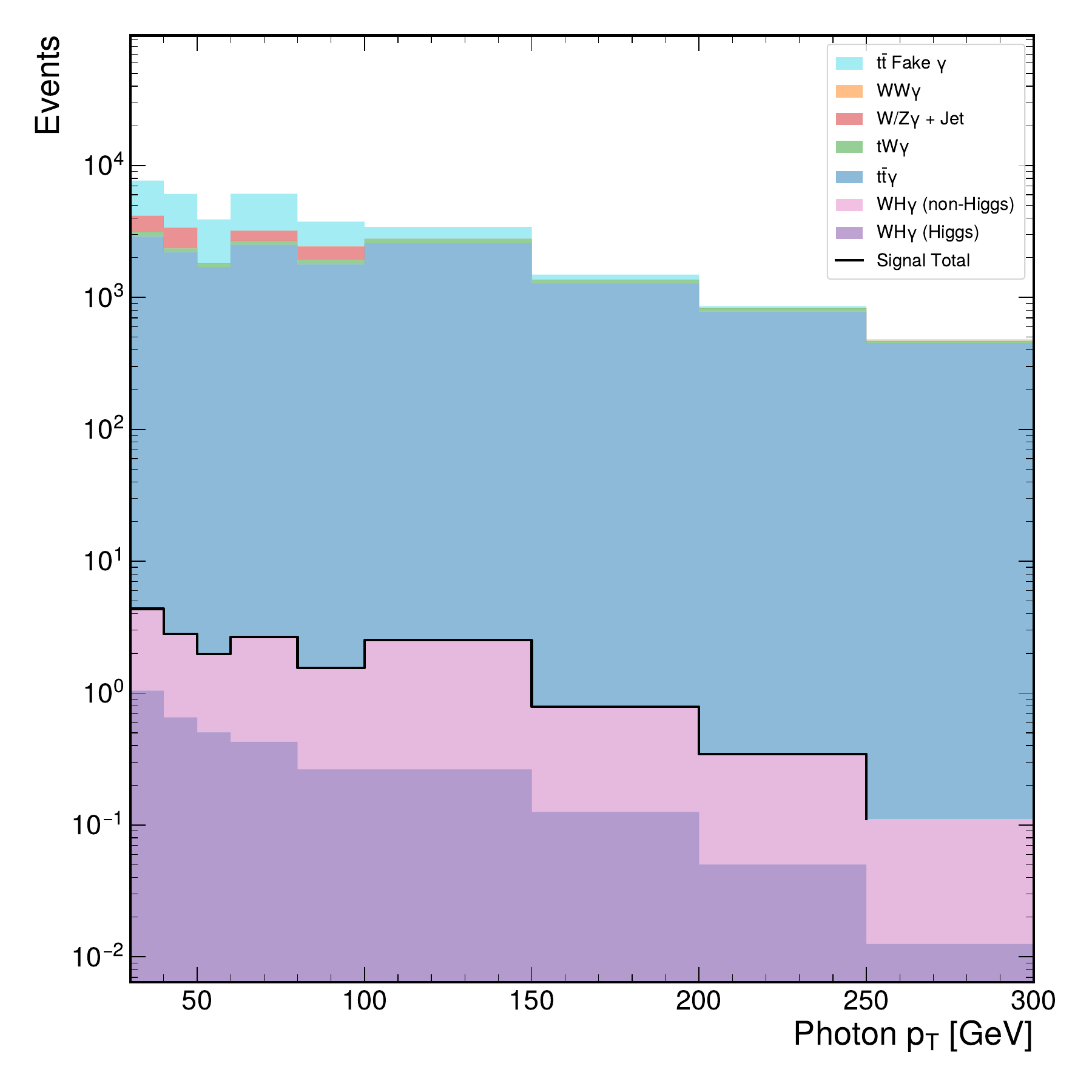}
    \includegraphics[width=0.48\textwidth]{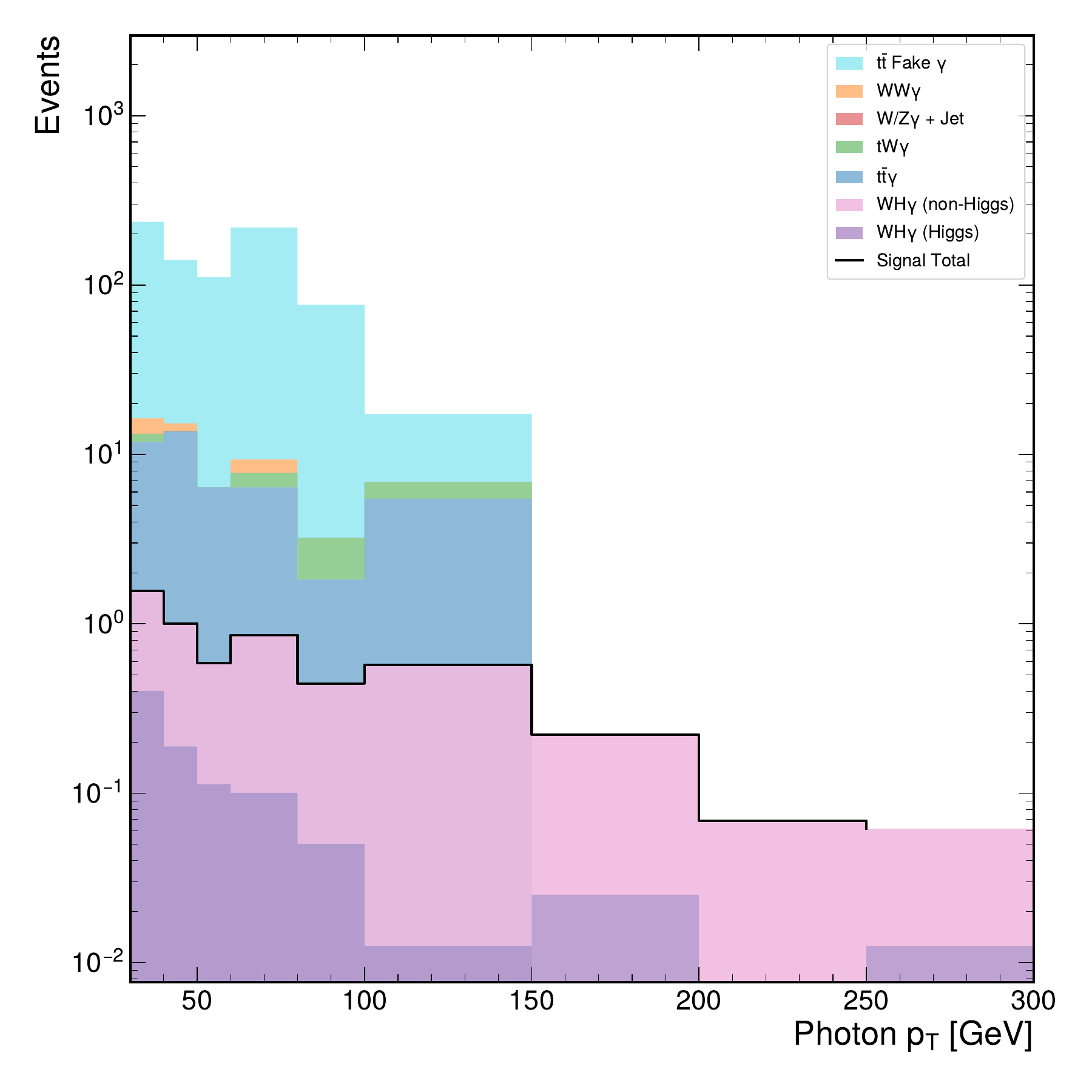}
    \caption{Left: Photon transverse momentum distribution before selection. Right: Photon transverse momentum distribution after selection. In Luminosity of 440 \(\rm fb^{-1}\).}
    \label{fig:fphotonpt}
\end{figure}

The fake photon background primarily originates from soft QCD processes where jets are misidentified as photons. We estimate this background using \Delphes simulation configured for the CMS detector. To ensure consistency with the $t\bar{t}\gamma$ signal definition and avoid phase space overlap, we adopt a specific truth-matching criterion: if a reconstructed photon matches a generator-level quark or gluon, it is classified as a true photon; otherwise, it is identified as a fake. However, the fidelity of this estimation is constrained by the limitations of fast simulation, which lacks key shower shape variables—such as $\sigma_{i\eta i\eta}$ and $H/E$—typically used for photon discrimination in experimental data. As illustrated in Figure~\ref{fig:fphotonpt}, the fake photon background dominates the low \pT region. Consequently, we impose a stringent photon \pT threshold to suppress this reducible background and focus the analysis on the irreducible component, deferring a detailed study of fake photons to future work.

\section{BDT Training and Result}

We have used the multivariate technique BDT (Boosted Decision Tree) to combined variable as BDT score. We choose \BDT model as a more general usage in machine learning region.

For pre-select, in order to maintain the generalization of the model, we need to keep more information when training the model, so we use a looser cut. Therefore, we require that transverse momentum \(\pTlep > 15 \GeV\) and pseudorapidity \(|\eta^{\ell}| < 2.4\) for both electron and muon. In this step the efficiency for signal process is near 5\%. We select series of kinematic variables as input of BDT input. Both electron and muon provide \pT, \(\eta\) and \(\phi\) as input variables. In addition, photon require energy number cluster in ECAL after it contain kinematic variables like leptons. The missing transverse momentum \pTmiss is also included as input variable. The efficiency of events for each process after pre-selection and BDT cut are listed in Table~\ref{tab:eff}.

For jets it could be different because \BDT need variables arranged by column. Therefore we need summary jet identity for each event. We choose jets kinematic information include \pT, \(\eta\), \(\phi\) and mass \(M\), and cluster identity \(\Delta \eta\) and \(\Delta \phi\), with flavour classify Btag as input variables. Then we re-arrange this variables in event level. For each variable \(v\), we calculated its characteristics include:
\begin{itemize}
    \item sum: \(\sum_i v_i\) 
    \item mean \(\bar{v}\): \(\frac{1}{n}\sum_i v_i\)
    \item standard error: \(\sum_i(v_i-\bar{v})^2\)
    \item count: \(i= 1, \dots, n\)
    \item leading number: largest \(v_i\) and next largest \(v_i\)
\end{itemize}
where \(i\) is index for all jets in events. For btag, we only need two index, count for number of btag jets and ratio for number of btag jets to total jets. In this way, we can keep the same number of input variables for each event. 

\begin{table}[htb]
    \centering
    \label{tab:eff}
    \caption{Efficiency after pre-selection training sample}
    \begin{tabularx}{0.5\textwidth}{l|X}
        \toprule
        Process & Efficiency \\
        \midrule
        signal(Higgs) & 1.4\% \\
        signal(Non-Higgs) & 7.6\% \\
        \(t\bar{t}\gamma\) & 12.2\% \\
        \(tW\gamma/\bar{t}W\gamma\) & 14.1\% \\
        \(W/Z+\gamma+\rm Jets\) & 5.4\% \\
        \(WW\gamma/W Z\gamma\) & 12.3\% \\
        \bottomrule
    \end{tabularx}
\end{table}

For training, we generate one hundred thousand events for each signal and background process. The signal events are assigned a label of 1, while the background events are labeled as 0. The training process is based on the \BDT Python package. We split the dataset into two parts: 80\% of the events are used for training the model, while the remaining 20\% are reserved for validation purposes after reshuffling. The signal region is defined by the output score of the BDT model to be greater than 0.5, and apply selection on BDT cut result.

\begin{figure}[htb]
    \centering
    \includegraphics[width=0.48\textwidth]{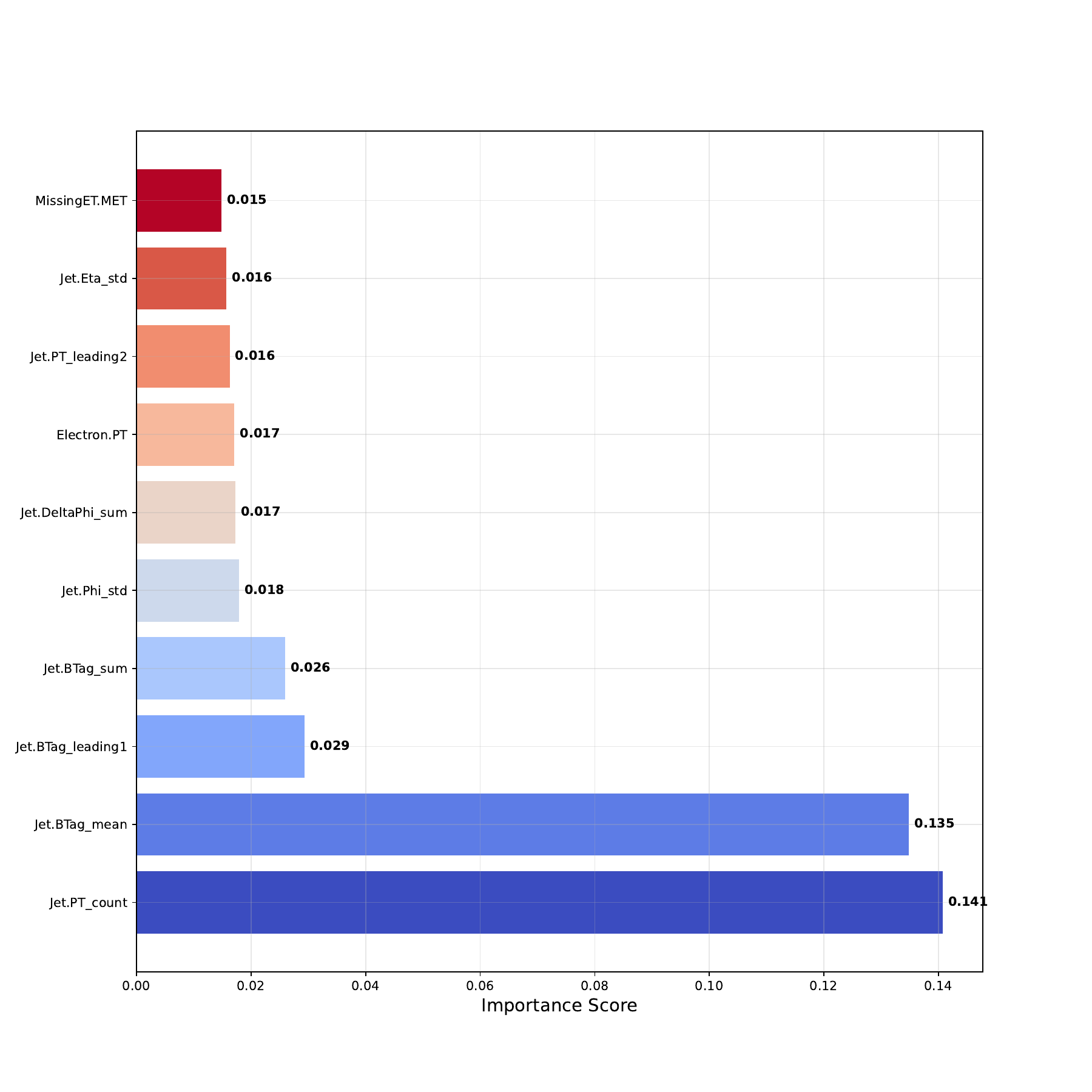}
    \includegraphics[width=0.48\textwidth]{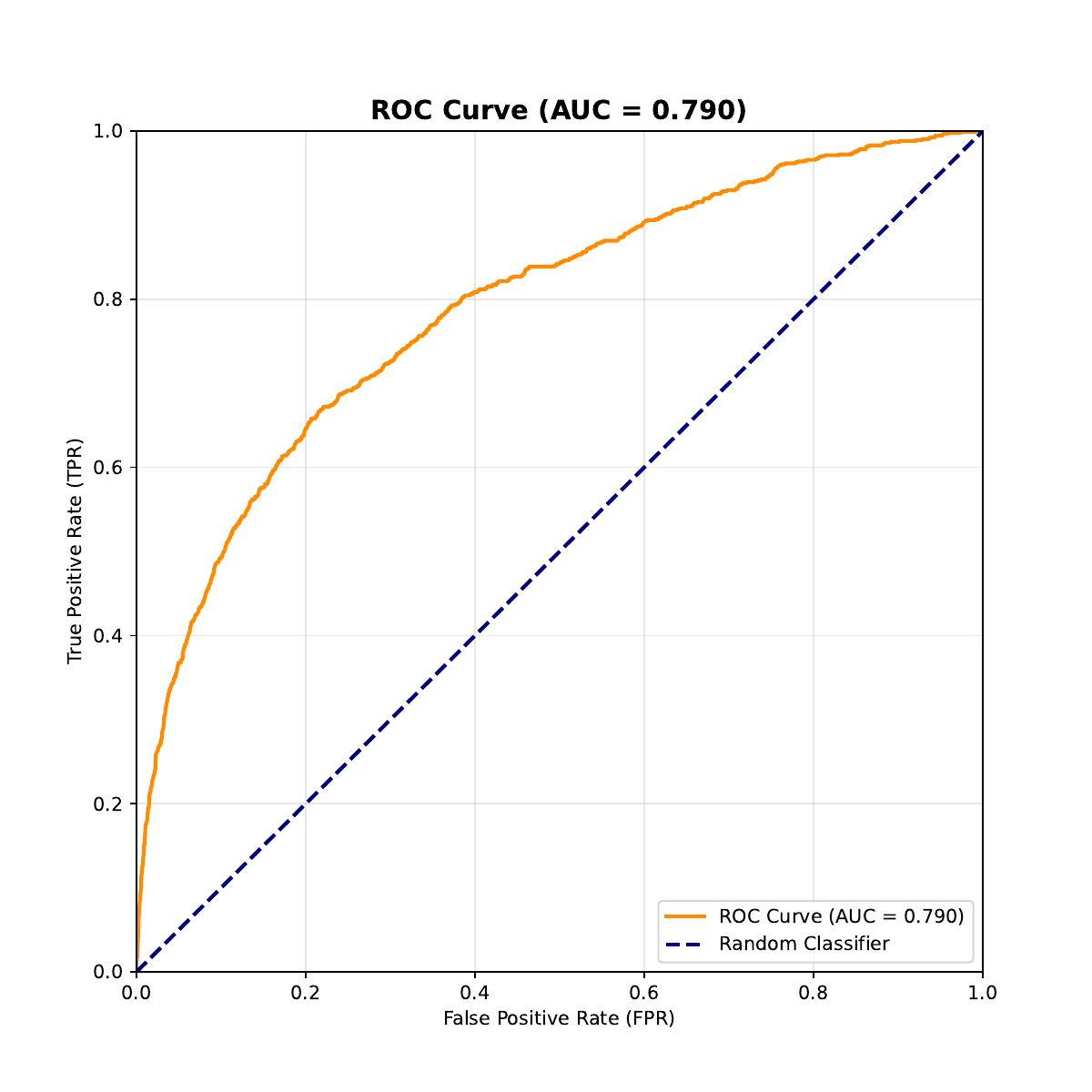}
    \caption{Left: Feature importance of BDT training (top 10 variables). Right: ROC curve of BDT training (AUC = 0.790).}
    \label{fig:bdt_results}
\end{figure}


\begin{figure}[htb]
    \centering
    \includegraphics[width=0.48\textwidth]{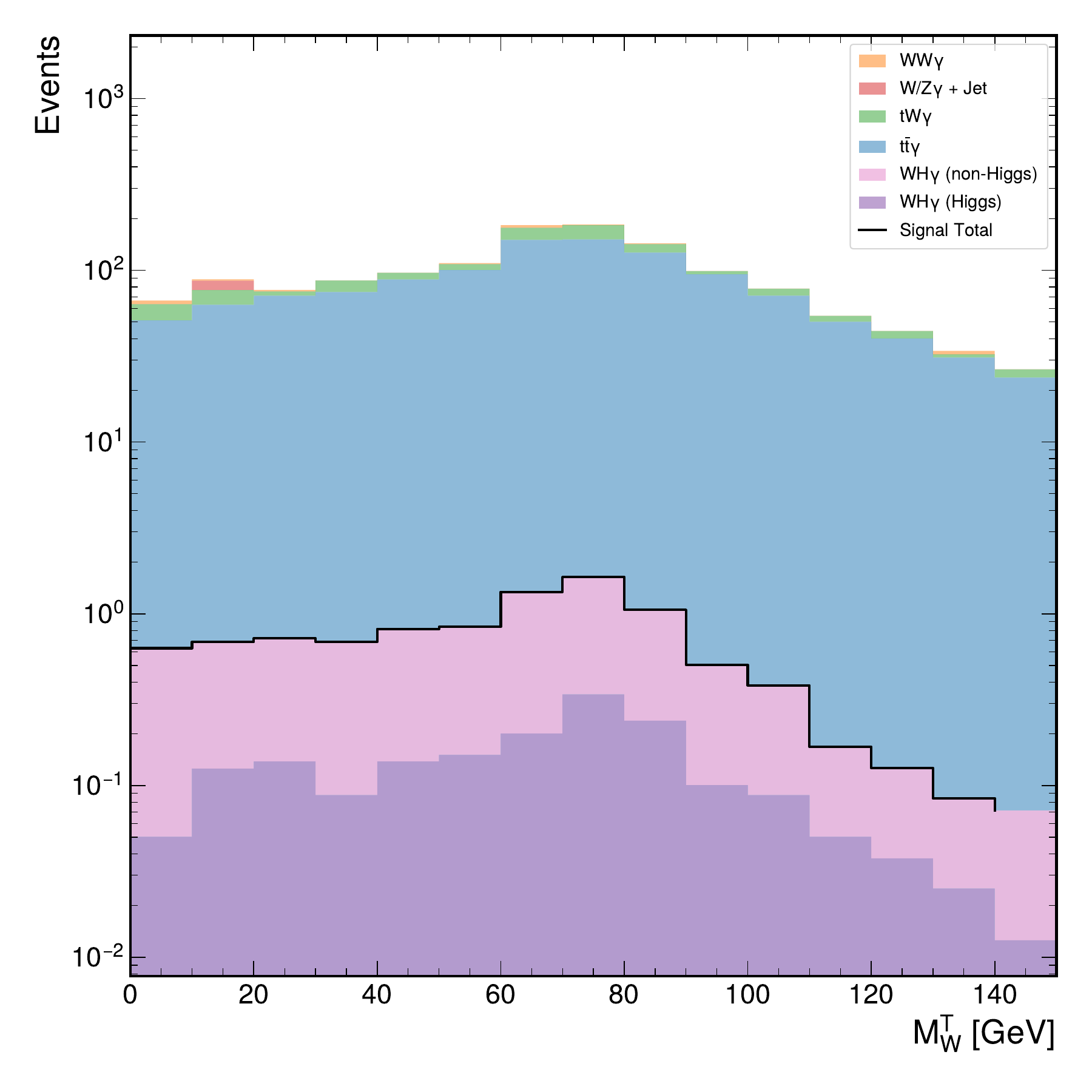}
    \includegraphics[width=0.48\textwidth]{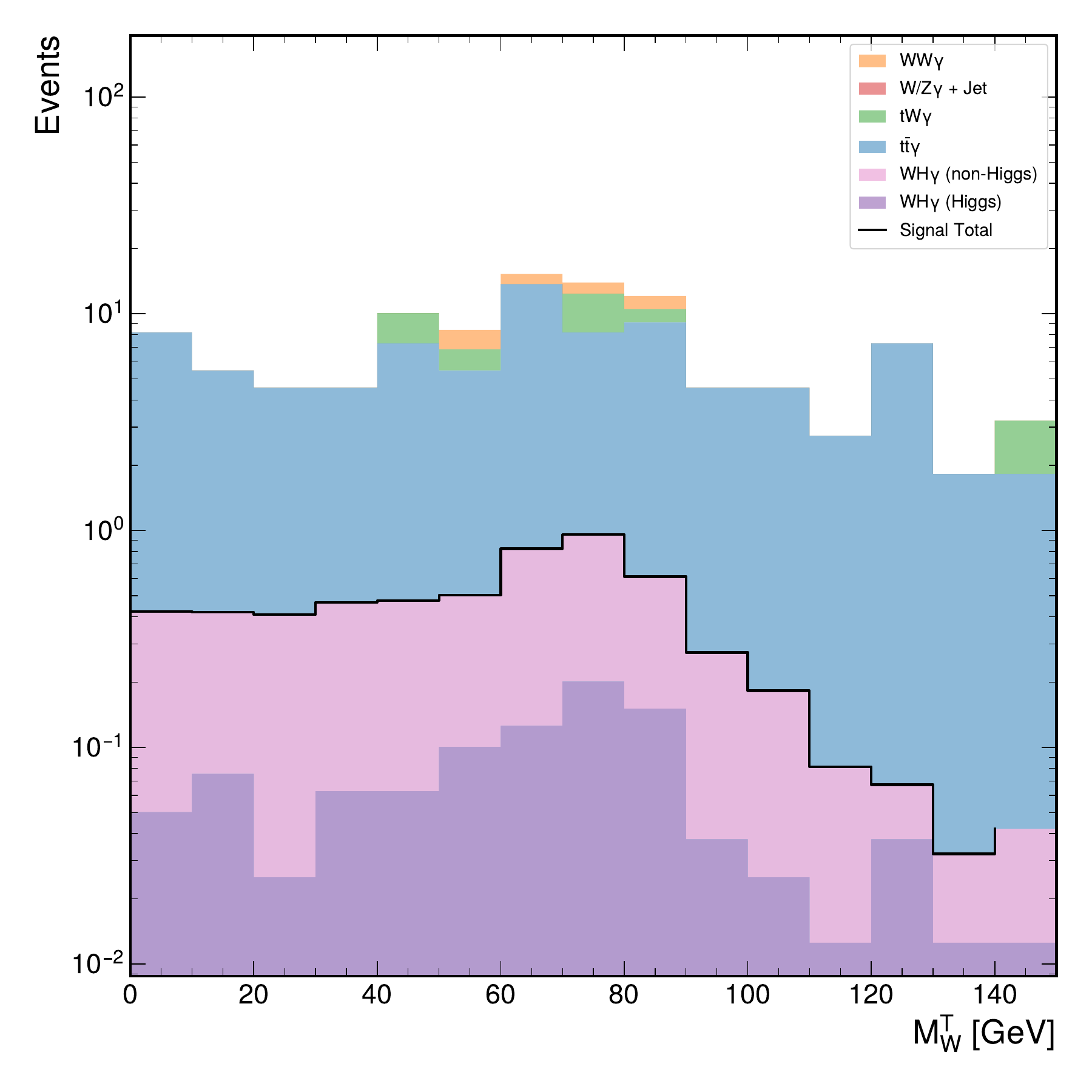}
    \caption{Left: Transverse W boson mass distribution without BDT cut. Right: Transverse W boson mass distribution with BDT cut.}
    \label{fig:BDTcutWmass}
\end{figure}

The transverse W boson mass is from lepton and missing transverse momentum as \(\mt ^{W} = \sqrt{2\ \pTlep \cdot \pTmiss \cdot(1-\cos\Delta \varphi)}\) where \(\Delta \varphi\) is the angle between lepton and missing transverse momentum in the transverse plane.

As illustrated in Fig.~\ref{fig:BDTcutWmass}, the application of the BDT selection leads to a substantial suppression of the background, reducing it by approximately two orders of magnitude, while the signal distribution remains essentially unchanged. This demonstrates the strong discriminating power of the BDT, which efficiently retains signal events and eliminates a significant fraction of background contamination. Such a pronounced reduction in background greatly enhances the statistical significance of the signal.

The BB mass is reconstructed from two leading b-tagged jets in the event by combining their four-momenta. 
\begin{figure}[htb]
    \centering
    \includegraphics[width=0.6\textwidth]{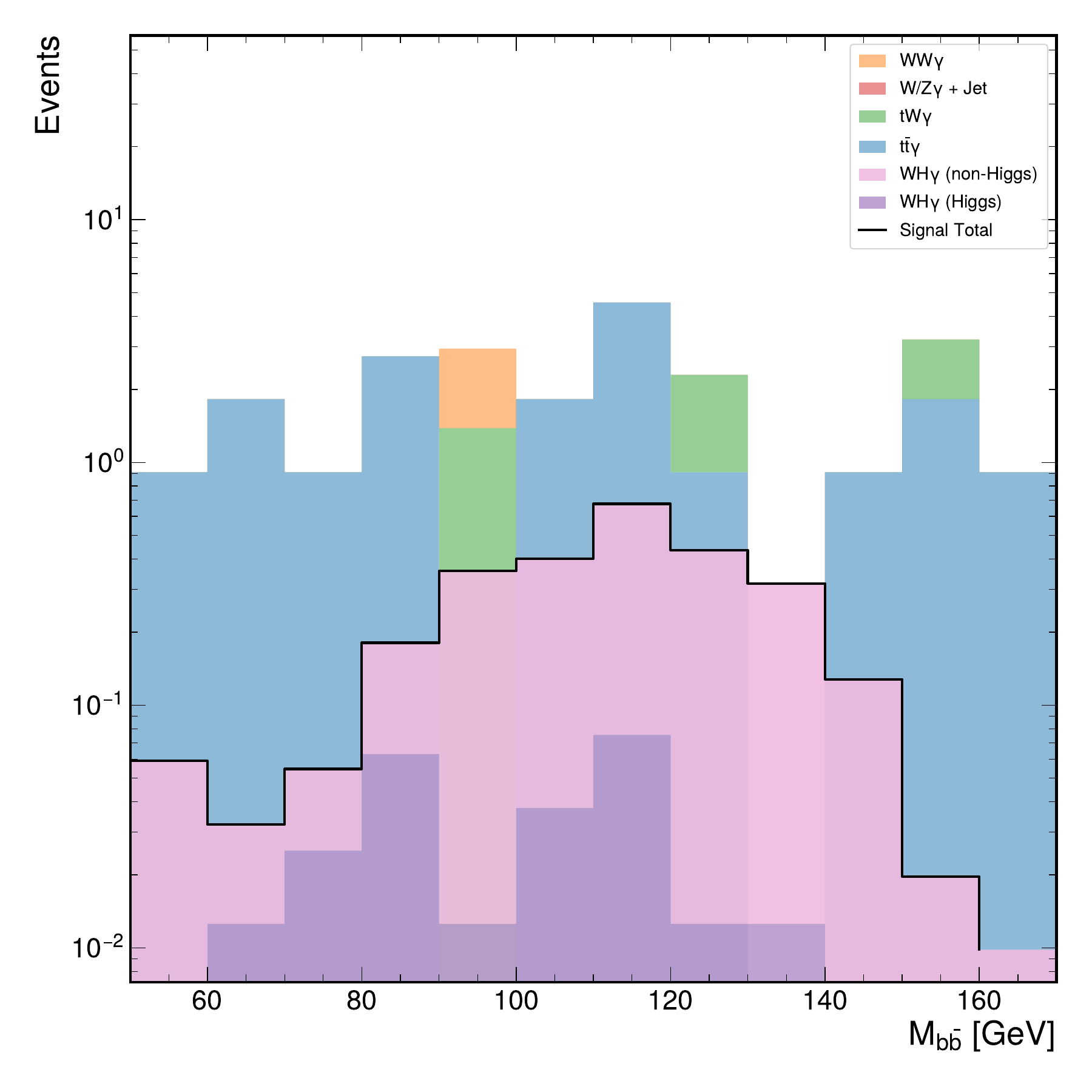}
    \caption{BB mass distribution with BDT cut, and selecting events with 2 or 3 jets and $\pT > 50 \textrm{GeV}$.} 
    \label{fig:BDTcutBBmass}
\end{figure}
Fig.~\ref{fig:BDTcutBBmass} shows the BB mass distribution after applying the BDT cut and selecting events with 2 or 3 jets. We can see a peak of \WHG (non-higgs) around 125 GeV, which is consistent with the Higgs boson mass because the two b-jets are from Higgs boson decay. The observed signal (Higgs) peak appears slightly below 125 GeV, which can be attributed to energy loss resulting from photon radiation off the Higgs boson.

We assume the luminosity of LHC is 440 \(\rm fb^{-1}\) as simulation of \texttt{CMS} experiment Run-3 + Run-2 . The significance \(\sigma\) is calculated using the formula \(S/\sqrt{B}\). By applying a BDT cut of 0.6, we obtain the event yields and significance listed in Table~\ref{tab:significance}. The results indicate that the significance of the \WHG (non-higgs) process is approximately four times greater than that of the \WHG (higgs) process. For LHC Run-3 + Run-2 with an integrated luminosity of 440 \(\rm fb^{-1}\), the combined significance of both processes reaches nearly 0.5\(\sigma\), and if we consider LHC HL-LHC (High Luminosity LHC) with an integrated luminosity of 3000 \(\rm fb^{-1}\), the significance could potentially approach 2\(\sigma\).

\begin{table}[htb]
    \centering
    \caption{Event significance after BDT cut at 13 TeV at \(M_{bb} \in [75, 150] \rm GeV\)}
    \label{tab:significance}
    \begin{tabularx}{\textwidth}{l|XXXX}
        \toprule
        Signal & Signal Event & Background Event & Significance \(\sigma\)\newline with 440 \(\rm fb^{-1}\) & Significance \(\sigma\)\newline with 3000 \(\rm fb^{-1}\)\\
        \midrule
        \WHG (higgs) & 0.226 & 16.190 & 0.056 & 0.147 \\
        \WHG (non-higgs) & 2.301 & 16.190 & 0.572 & 1.494 \\
        \WHG Sum & 2.528 & 16.190 & 0.628 & 1.640 \\
        \bottomrule
    \end{tabularx}
\end{table}

\section{Summary}

We present a projected study of \WHG production at the CMS experiment, focusing on the final state where the W boson decays leptonically (into an electron or muon and a neutrino) and the Higgs boson decays into a $b\bar{b}$ pair. This rare triboson process serves as a simultaneous probe of the Standard Model (SM) electroweak sector, making it sensitive to potential new physics affecting electroweak couplings. Based on the BDT analysis summarized in Table~\ref{tab:significance}, the expected significance with an integrated luminosity of 440~$\rm fb^{-1}$ is approximately 0.63$\sigma$. With the full High-Luminosity LHC (HL-LHC) dataset of 3000~$\rm fb^{-1}$, the significance is projected to reach 1.64$\sigma$. While the observation of this process remains challenging, these projections motivate the initiation of an experimental search to constrain the relevant phase space.


\begin{acknowledgments}
This work is supported in part by the National Natural Science Foundation of China under Grants No. 12325504 and No. 12061141002.
\end{acknowledgments}


\bibliographystyle{JHEP}
\bibliography{biblio.bib}

\end{document}